\documentclass[12pt]{article}
\usepackage[cp1251]{inputenc}
\usepackage{amsfonts}
\usepackage[english]{babel}
\usepackage{eufrak}
\usepackage{setstack}
\usepackage{cite}
\usepackage{amssymb}
\title{ p-adic Gauss integrals from the Poison summarizing formula}
\begin{document}
\author{D.V. Prokhorenko \footnote{Institute of Spectroscopy, Russian Academy of Sciences, 142190 Moskow Region, Troitsk,  prokhordv@yandex.ru}}

 \maketitle
\begin{abstract}
In the present paper we show how to obtain the well-known formula
for Gauss sums and the Gauss reciprocity low from the Poison
summarizing formula by using some ideas of renormalization and
ergodic theories. We also apply our method to obtain new simple
derivation of the standard formula for \(p\)-adic Gauss integrals.
\end{abstract}
\newpage
\section{Notations}
Let \(S(R)\) be a Schwartz space on real line (the space of all
smooth functions decaying at infinity faster than each inverse
polynomial with all their derivatives). Let \(f(x) \in S(R)\) be
some function from the Schwartz space. The Fourier transform of
\(f(x)\) is defined by the following formula:
\begin{eqnarray}
\hat{f}(k)=(2 \pi)^{-\frac{1}{2}}\int f(x)e^{ikx}dx
\end{eqnarray}
The inverse Fourier transform is defined as follows
\begin{eqnarray}
 \check{f}(x)=(2 \pi)^{-\frac{1}{2}}\int f(k)e^{-ikx}dk
\end{eqnarray}
It is easy to prove that the Fourier transform ant its inverse map
the Schwartz space into the Schwartz space. The Fourier theorem
states that \(\forall f(x) \in S(R)\)
\begin{eqnarray}
\check{\hat{f}}(x)=\hat{\check{f}}(x)=f(x).
\end{eqnarray}
Let \(f(x) \in S(R)\). We have
\begin{eqnarray}
\sum \limits_{k \in \mathbb{Z}}\hat{f}(k)=(2 \pi)^{-\frac{1}{2}}\sum
\limits_{k \in \mathbb{Z}}\int f(x)e^{ikx}dx= \sum
\limits_{k\in\mathbb{Z}} (2\pi)^{-\frac{1}{2}} \sum \limits_{m \in
\mathbb{Z}}\int\limits_{2\pi(m-1/2)}^{2\pi(m+1/2)} f(x)e^{ikx}dx\nonumber\\
=\sum \limits_{m \in\mathbb{Z}} (2\pi)^{-\frac{1}{2}} \sum
\limits_{k \in \mathbb{Z}}\int \limits_{2\pi(m-1/2)}^{2\pi(m+1/2)}
f(x)e^{ikx}dx =(2\pi)^{\frac{1}{2}} \sum \limits_{m \in
\mathbb{Z}}f(2\pi m).
\end{eqnarray}
Therefore we have the following Poison summarizing formula
\begin{eqnarray}
\sum \limits_{k \in \mathbb{Z}}\hat{f}(k)=(2\pi)^{\frac{1}{2}} \sum
\limits_{m \in \mathbb{Z}}f(2\pi m).
\end{eqnarray}
The rigorous proof of this formula see for example in [1, 2].

Let \(p\) be an integer primer number. The Gauss sums are defined as
follows:
\begin{eqnarray}
G.S.(\frac{q}{p}):=\sum \limits_{k=0}^{p-1} {\rm exp \mit}(2\pi i
\frac{q}{p} k^2).
\end{eqnarray}
Let us introduce the Legendre symbols \((\frac{p}{q})\). By
definition
\begin{eqnarray}
(\frac{q}{p})=1\; \rm if \mit\; \exists\; \rm integer \mit
\;x:\;\mit q=x^2 (\rm mod \mit\;p)
\end{eqnarray}
and
\begin{eqnarray}
(\frac{q}{p})=-1\; \rm if \mit\; \forall\; \rm integer \mit
\;x:\;\mit q\neq x^2 (\rm mod \mit\;p)
\end{eqnarray}
The integer numbers \(q\) such that \((\frac{q}{p})=1\) are called
quadratic residues, and the integer numbers such that
\((\frac{q}{p})=-1\) are called quadratic nonresidues.
 Note that
\begin{eqnarray}
\sum \limits_{k=0}^{p-1} {\rm exp \mit}(2 \pi i k)=0
\end{eqnarray}
Note also that the number of quadratic residues is equal to the
number of quadratic nonresidues. Therefore, we have
\begin{eqnarray}
G.S.(\frac{q}{p})=(\frac{q}{p})G.S.(\frac{1}{p}).
\end{eqnarray}
Let \(\alpha \in \mathbb{C}\) and \(\mathfrak{R} \alpha >0\). The
Gauss integral, by definition, is the following integral
\begin{eqnarray}
\int e^{-\alpha x^2} dx=\sqrt{\frac{\pi}{\alpha}}, \label{Gauss}
\end{eqnarray}
where we take such branch of square root that it \(>0\) if
\(\alpha\) real and \(\alpha>0\).

One can proof (by using (\ref{Gauss}) that
\begin{eqnarray}
\widehat{e^{-\alpha
x^2}}(k)=\frac{1}{\sqrt{2\alpha}}e^{-\frac{1}{4}\frac{k^2}{\alpha}}.
\end{eqnarray}
\section{The Gauss sums and the Gauss reciprocity low}
Instead of calculation
\begin{eqnarray}
G.S.(\frac{1}{p}):=\sum \limits_{k=0}^{p-1} {\rm exp \mit}(2\pi i
\frac{1}{p} k^2)
\end{eqnarray}
we will calculate

\begin{eqnarray}
G.S.^\varepsilon(\frac{1}{p}):=\sum \limits_{k\in \mathbb{Z}} {\rm
exp \mit}( \frac{2\pi i k^2}{p} -\varepsilon k^2), \nonumber\\
\varepsilon>0
\end{eqnarray}
Let us clear the question how \(G.S.^\varepsilon(\frac{1}{p})\)
connects with \(G.S.(\frac{1}{p})\).

For small \(\varepsilon>0\) we can regard the function
\(e^{-\varepsilon x^2}\) as constant function at each interval
\([mp,(m+1)p]\) for each integer \(m\). So, the part of
\(G.S.^\varepsilon(\frac{1}{p})\) calculated for
\(k=mp,\;mp+1,...(m+1)p-1\) can be approximated by
\(e^{-\varepsilon(mp)^2}G.S.(\frac{1}{p})\). Obtained sum over
points \(mp,\;m \in \mathbb{Z}\) we can approximate by integral. The
mistake will be of order \(O(1)\), \(\varepsilon\rightarrow 0\).
Therefore
\begin{eqnarray}
G.S.^\varepsilon(\frac{1}{p})=G.S.(\frac{1}{p})\sum \limits_{x \in
p\mathbb{Z}}{\rm exp \mit}(-\varepsilon x^2)+O(1)\nonumber\\
=G.S.(\frac{1}{p})\frac{1}{p}\int \limits_{-\infty}^{+\infty} \rm
exp \mit
(-\varepsilon x^2)dx +O(1)\nonumber\\
=G.S.(\frac{1}{p})\frac{1}{p}\sqrt{\frac{\pi}{\varepsilon}}+O(1),
\;\varepsilon\rightarrow 0.
\end{eqnarray}
Therefore we have the following representation
\begin{eqnarray}
G.S.^\varepsilon(\frac{1}{p})=G.S.(\frac{1}{p})\frac{1}{p}\sqrt{\frac{\pi}{\varepsilon}}+O(1),
\;\varepsilon\rightarrow 0. \label{I}
\end{eqnarray}
From other hand, let us denote by \(f(x)\) the following function:
\begin{eqnarray}
f(x):=\rm exp \mit (2\pi i \frac{x^2}{p}-\varepsilon x^2).
\end{eqnarray}
Let us calculate the Fourier transform of \(f(x)\). We have
\begin{eqnarray}
\hat{f}(k)=\frac{1}{\sqrt{2(\varepsilon-\frac{2\pi i}{p})}} \rm exp
\mit (-\frac{k^2}{4(\varepsilon-\frac{2\pi
i}{p})})\nonumber\\
\approx\frac{1}{\sqrt{-\frac{4\pi i}{p}}} \rm exp \mit (-\frac{k^2
i}{4(\frac{2\pi }{p})}(1+\frac{p\varepsilon}{2 \pi i}))
\end{eqnarray}
In result, we have
\begin{eqnarray}
\hat{f}(k)\approx \sqrt{\frac{p}{4\pi}}e^{i\frac{\pi}{4}}\rm exp
\mit (-\frac{k^2i}{4(\frac{2\pi}{p})}(1+\frac{p\varepsilon}{2 \pi
i})).
\end{eqnarray}
Using Poison summarizing formula we find:
\begin{eqnarray}
G.S.^\varepsilon(\frac{1}{p})=(2\pi)^{\frac{1}{2}} \sum \limits_{k
\in\mathbb{Z}} \hat{f}(2\pi k)\nonumber\\
=(2\pi)^{\frac{1}{2}}\sqrt{\frac{p}{4\pi}}e^{\frac{i\pi}{4}} \sum
\limits_{k=1}^4 \rm exp \mit (-\frac{2\pi i k^2}{4}p)\times
\frac{1}{4} \int \limits_{-\infty}^{+\infty} {\rm exp \mit}
(-\frac{k^2p^2\varepsilon}{4})+O(1)\nonumber\\
=(2\pi)^{\frac{1}{2}}\frac{e^{\frac{i\pi}{4}}}{8}\sqrt{\frac{p}{\pi}}
\sqrt{\frac{4\pi}{\varepsilon}}\frac{1}{p}\sum \limits_{k=1}^4 \rm
exp \mit (-\frac{2\pi i k^2}{4}p)+O(1),\;\varepsilon\rightarrow 0.
\end{eqnarray}
In result we have:
\begin{eqnarray}
G.S.^\varepsilon(\frac{1}{p})=\frac{e^{\frac{i\pi}{4}}}{2\sqrt{2}}\frac{1}{\sqrt{p}}(\frac{\pi}{\varepsilon})^{\frac{1}{2}}
\sum \limits_{k=1}^{4} \rm exp \mit (-\frac{2\pi i
k^2}{4}p)+O(1),\;\varepsilon\rightarrow 0. \label{II}
\end{eqnarray}
So we need to calculate the sum in right hand side of last equality.
Squares of numbers \(1,\;2,\;3,\;4\) are \(0,\;1,\;0,\;1\) modulo 4
respectively. So
\begin{eqnarray}
\sum \limits_{k=1}^4 \rm exp \mit (-\frac{2\pi i
k^2}{4}p)=2\sqrt{2}e^{-\frac{i\pi}{4}},\; {\rm if \mit}\; p=1
\,({\rm
mod \mit}\, 4), \nonumber\\
\sum \limits_{k=1}^4 \rm exp \mit (-\frac{2\pi i
k^2}{4}p)=2\sqrt{2}e^{\frac{i\pi}{4}},\; {\rm if \mit}\; p=3\, ({\rm
mod \mit}\, 4),
\end{eqnarray}
Therefore from (\ref{I}) and (\ref{II}) we have
\begin{eqnarray}
G.S.(\frac{q}{p})=\sqrt{p} (\frac{q}{p}),\; {\rm if \mit}\;
p=1\,(\rm mod\mit\, 4),\nonumber\\
G.S.(\frac{q}{p})=i \sqrt{p} (\frac{q}{p}),\; {\rm if \mit}\;
p=3\,(\rm mod\mit\, 4),
\end{eqnarray}
that has to be proven.

Let us now use this method to prove the Gauss reciprocity low. Put
by definition
\begin{eqnarray}
G.S.^{\varepsilon}(\frac{q}{p}):=\sum \limits_{x \in \mathbb{Z}}\rm
exp \mit (\frac{2\pi i q x^2}{p}-\varepsilon x^2) \label{III}
\end{eqnarray}
and
\begin{eqnarray}
f_{\frac{q}{p}}(x):= \rm exp \mit (\frac{2 \pi i
q}{p}x^2-\varepsilon x^2)
\end{eqnarray}
By the same method as we prove (\ref{I}) it is easy to prove that
\begin{eqnarray}
G.S.^{\varepsilon}(\frac{q}{p})=G.S.(\frac{q}{p})\frac{1}{p}\sqrt{\frac{\pi}{\varepsilon}}+O(1),\;
\varepsilon \rightarrow 0. \label{IV}
\end{eqnarray}
It is easy also to calculate that
\begin{eqnarray}
\hat{f}_{\frac{q}{p}}(k)\approx \sqrt{\frac{p}{4\pi
q}}e^{\frac{i\pi}{4}}\rm exp \mit (-\frac{i k^2p}{4(2\pi
q)}(1+\frac{p\varepsilon}{i2\pi q}))
\end{eqnarray}
Let us now apply to (\ref{III}) the method used for derivation
(\ref{II}). We obtain
\begin{eqnarray}
G.S.^{\varepsilon}(\frac{q}{p})=(2\pi)^{\frac{1}{2}}\sum \limits_{k
\in \mathbb{Z}}\hat{f}_{\frac{q}{p}}(2\pi k)\nonumber\\
=(2\pi)^{\frac{1}{2}}\sqrt{\frac{p}{4\pi
q}}e^{i\frac{\pi}{4}}\frac{1}{4} \int \limits_{-\infty}^{+\infty}
\rm exp \mit(-\frac{k^2p^2}{4}\varepsilon)dp\nonumber\\
\times \sum \limits_{k=1}^{4q} \rm exp \mit (-\frac{2\pi
i}{4}k^2\frac{p}{q})+O(1),\;\varepsilon\rightarrow 0.
\end{eqnarray}
In result
\begin{eqnarray}
G.S.^{\varepsilon}(\frac{q}{p})=\frac{e^{i\frac{\pi}{4}}}{2\sqrt{2}}\frac{1}{\sqrt{pq}}(\frac{\pi}{\varepsilon})^{\frac{1}{2}}\sum
\limits_{k=1}^{4q} \rm exp \mit (-\frac{2\pi i}{4} k^2
\frac{p}{q})+O(1),\;\varepsilon\rightarrow 0. \label{V}
\end{eqnarray}
So we need to calculate:
\begin{eqnarray}
\Sigma:=\sum \limits_{k=1}^{4q} \rm exp \mit (-\frac{2\pi
i}{4}k^2\frac{p}{q}).
\end{eqnarray}
Let \(a\) and \(b\) be integer reciprocity primer numbers. There
exists a ring homomorphism
\begin{eqnarray}
\mathbb{Z}_{ab}\rightarrow \mathbb{Z}_a\times \mathbb{Z}_b
\end{eqnarray}
which assign to each element \(x(\rm mod \mit\, ab) \in
\mathbb{Z}_{ab}\) the element \((x(\rm mod\mit\, a),x(\rm mod\mit\,
b))\in \mathbb{Z}_a\times \mathbb{Z}_b\)
\begin{eqnarray}
x(\rm mod \mit\, ab) \mapsto(x(\rm mod\mit\, a),x(\rm mod\mit\, b))
\end{eqnarray}
Well known theorem (Chinese remainder theorem) states that the
homomorphism just described is an isomorphism.

\(a\) and \(b\) are reciprocity primer numbers. So there exists such
integers \(r\) and \(s\) that \(ra+sb=1\). Therefore if \((x,y)\in
\mathbb{Z}_a \times \mathbb{Z}_b\) then its inverse image under the
homomorphism just defined is
\begin{eqnarray}
sbx+ray.
\end{eqnarray}
If the element \(\alpha \in \mathbb{Z}_{ab}\) is a square (i.e.
\(\alpha=\beta^2\) for some integer \(\beta\)) then its image
\((x,y)\) is also a square i.e. there exists integers \((z,w)\) such
that \(x=z^2\),\(y=w^2\). It follows from this remark that if
\(q\neq 2\) then
\begin{eqnarray}
\Sigma =\sum \limits_{k=1}^{4q} \rm exp \mit (-\frac{2\pi
i}{4}k^2\frac{p}{q})\nonumber\\
=\sum \limits_{n=1}^{4} \rm exp \mit (-\frac{2\pi i}{4}spn^2)\sum
\limits_{m=1}^q(-\frac{2\pi i}{q}r p m^2),
\end{eqnarray}
where \(s\) and \(r\) are integer satisfying to
\begin{eqnarray}
4r+qs=1.
\end{eqnarray}
Let us calculate \(r\) and \(s\). Consider the case \(q=1(\rm mod
\mit\,4)\), i.e. \(q=4f+1\) for some integer \(f\). In this case
\(s=1\) and \(r=-f=-\frac{-1+q}{4} (\rm mod \mit\, q)=\frac{1}{4}
(\rm mod \mit \,q)\). Consider the case \(q=4f-1\) i.e. \(q=-1(\rm
mod \mit\,4)\). In this case \(s=-1\) and \(r=f=\frac{1+q}{4} (\rm
mod \mit\, q)=\frac{1}{4} (\rm mod \mit\, q)\).

Let us calculate \(\Sigma\) for the case \(q=4f+1\) for some integer
\(f\). In this case
\begin{eqnarray}
\Sigma=2\sqrt{2}e^{-i\frac{\pi}{4}}(-1)^{\frac{q-1}{2}}G.S.(\frac{p}{q})\;
\rm if \mit\; p=1(\rm mod
\mit \,4)\nonumber\\
\Sigma=2\sqrt{2}e^{i\frac{\pi}{4}}(-1)^{\frac{q-1}{2}}G.S.(\frac{p}{q})\;
\rm if \mit\; p=3(\rm mod \mit\,4).
\end{eqnarray}

Consider now the case \(q=4f-1\) for some integer \(f\). In this
case we have
\begin{eqnarray}
\Sigma=2\sqrt{2}e^{i\frac{\pi}{4}}(-1)^{\frac{q-1}{2}}G.S.(\frac{p}{q})\;
\rm if \mit\; p=1(\rm mod \mit\,4)\nonumber\\
\Sigma=2\sqrt{2}e^{-i\frac{\pi}{4}}(-1)^{\frac{q-1}{2}}G.S.(\frac{p}{q})\;
\rm if \mit\; p=3(\rm mod \mit\,4).
\end{eqnarray}
In result (\ref{IV}) and (\ref{V}) implies
\begin{eqnarray}
G.S.(\frac{q}{p})\frac{1}{\sqrt{p}}=G.S.(\frac{p}{q})\frac{1}{\sqrt{q}}\alpha(p,q).\label{VI}
\end{eqnarray}
Here by definition
\(\alpha(p,q)=(-1)^{\frac{q-1}{2}\frac{p-1}{2}}\beta(p,q)\), where
\begin{eqnarray}
\beta(p,q)=1\; \rm if \mit\; p=1(\rm mod \mit)\, 4,\;\rm and \mit\;
q=1\;(\rm mod \mit)\, 4\nonumber\\
\beta(p,q)=-i\; \rm if \mit\; p=1(\rm mod \mit)\, 4,\;\rm and \mit\;
q=3\;(\rm mod \mit)\, 4\nonumber\\
\beta(p,q)=i\; \rm if \mit\; p=3(\rm mod \mit)\, 4,\;\rm and \mit\;
q=1\;(\rm mod \mit)\, 4\nonumber\\
\beta(p,q)=1\; \rm if \mit\; p=3(\rm mod \mit)\, 4,\;\rm and \mit\;
q=3\;(\rm mod \mit)\, 4
\end{eqnarray}
And by using the expression of Gauss sums trough the Legendre
symbols we obtain
\begin{eqnarray}
(\frac{q}{p})(\frac{p}{q})=(-1)^{\frac{p-1}{2}\frac{q-1}{2}},
\label{VII}
\end{eqnarray}
what has to be proven. The equality (\ref{VII}) is named the Gauss
reciprocity low.

\section{\(p\)-adic Gauss integrals}
Let \(p\) --- be an integer primer number. Let \(\mathbb{Q}\) be a
field of rational number. Each \(r \in \mathbb{Q}:\;r\neq 0\) can be
represented as follows:
\begin{eqnarray}
r=\frac{a}{b}p^\gamma,
\end{eqnarray}
where \(\gamma \in \mathbb{Z}\), \(a,b\in\mathbb{Z}\), \(a\neq0(\rm
mod \mit\;p)\), \(a\neq0(\rm mod \mit\;p)\). The field
\(\mathbb{Q}_p\) of \(p\)-adic numbers by definition is a completion
of \(\mathbb{Q}\) with respect to the following norm:
\begin{eqnarray}
|a|_p=|\frac{a}{b}p^\gamma|_p=p^{-\gamma}.
\end{eqnarray}
It is easy to prove that \(|\cdot|_p\) satisfy to the following
ultrametric property:
\begin{eqnarray}
|a+b|_p\leq \rm max \mit (|a|_p,|b|_p). \label{U}
\end{eqnarray}
Each \(a \in \mathbb{Q}_p\) can be represented as follows:
\begin{eqnarray}
a=p^{-\gamma}(a_0+a_1p+a_2p^2+...),
\end{eqnarray}
where \(\gamma \in \mathbb{Z}\), \(a_0,a_1,a_2\in \{0,1,...,p-1\}\)
and the series converges with respect to the \(p\)-adic norm
\(|\cdot|_p\).

Let \(a=p^{-\gamma}(a_0+a_1p+a_2p^2+...)\in\mathbb{Q}_p\), \(\gamma
\in \mathbb{Z}\), \(a_0,a_1,a_2\in \{0,1,...,p-1\}\). By definition
the fractional part \(\{a\}\) of \(a\) is defined as follows:
\begin{eqnarray}
\{a\}=p^{-\gamma}(a_0+a_1p+...+a_{\gamma-1}p^{\gamma-1})\;\rm
if\mit\;\gamma>0,\nonumber\\
\{a\}=0\;\rm if\mit\;\gamma\leq 0.
\end{eqnarray}
For each \(x\in\mathbb{Q}_p\) put by definition:
\begin{eqnarray}
\chi_p(x)=\rm exp \mit (2 \pi i \{x\}).
\end{eqnarray}
Let \(B_0=\{x \in \mathbb{Q}_p||x|_p\leq1\}\) be an unit ball in
\(\mathbb{Q}_p\) with the center at zero. Note that the ultrametric
condition (\ref{U}) implies that each point of unit ball \(B_0\) is
its center. \(B_0\) is a compact grope with respect to "\(+\)" as a
group operation. So there exists a Haar measure \(d\mu\) at
\(\mathbb{Q}_p\) which we will normalize such that \(\mu(B_0)=1\).

More about \(p\)-adic analysis see in [3].

Let \(\mathbb{R}^+=\{x\in\mathbb{R}|x\geq 0\}\). Let \(\Omega(x)\)
be a function \(\Omega: \mathbb{R}^+\rightarrow\mathbb{R}\) such
that \(\Omega(x)=1\; \rm if \mit\,x\leq 1\) and \(\Omega(x)=0\)
otherwise.

\textbf{Theorem.} Let \(p\) be a primer integer number such that
\(p\neq 2\). Then
\begin{eqnarray}
I:=\int \limits_{B_0} \chi_p(ax^2+bx)d\mu(x)=\Omega(|b|_p),\; \rm if
\;
\mit\; |a|_p\leq 1 \nonumber\\
I=\int \limits_{B_0}
\chi_p(ax^2+bx)d\mu(x)=\lambda_p(a)|a|_p^{-\frac{1}{2}}\chi_p(-\frac{b^2}{4a})\Omega
(|\frac{b}{a}|_p)\; \rm if \; \mit\; |a|_p> 1,
\end{eqnarray}
where
\begin{eqnarray}
\lambda_p(a)=1\;\rm if \mit\;|a|_p=p^N,\;N-\rm even, \mit \nonumber\\
\lambda_p(a)=(\frac{a_0}{p})
 \;\rm if \mit\;N-\rm odd \mit\;\rm and \mit\; p=1\;(\rm
 mod\mit\; 4), \nonumber\\
 \lambda_p(a)=i(\frac{a_0}{p})
 \;\rm if \mit\;N-\rm odd\;\rm and \mit \; p=3\;(\rm
 mod\mit\; 4)
\end{eqnarray}
 \textbf{Proof.}

a) Let \(|a|_p\leq 1\). In this case
\begin{eqnarray}
I=\int \limits_{B_0}\chi_p(ax^2+bx)d\mu(x)=\int
\limits_{B_0}\chi_p(bx)d\mu(x)=\Omega(|b|_p).
\end{eqnarray}

b) Now let \(|a|_p > 1\). If \(\left|\begin{array}{rcl}
\frac{b}{a}\end{array}\right|_p > 1\) then
\begin{eqnarray}
I=\int \limits_{B_0}\chi_p(ax^2+bx)d\mu(x)=\chi_p
\left(\begin{array}{rcl}
 -\frac{b^2}{4a} \end{array}\right)\int \limits_{B_0} \chi_p
\left(\begin{array}{rcl}a\left(\begin{array}{rcl}x^2+\frac{b}{a}x+\frac{b^2}{4a}
\end{array}\right)\end{array}\right)d\mu(x)\nonumber\\
\chi_p \left(\begin{array}{rcl}
 -\frac{b^2}{2a} \end{array}\right)\int \limits_{B_0} \chi_p
\left(\begin{array}{rcl}a\left(\begin{array}{rcl}x+\frac{b}{4a}
\end{array}\right)^2\end{array}\right)d\mu(x) \label{GI}
\end{eqnarray}
Put
\begin{eqnarray}
c:=\frac{b}{2a}=p^{-N}(c_0+c_1p+...),
\end{eqnarray}
and
\begin{eqnarray}
a=p^{-\gamma}(a_0+a_1p+...),
\end{eqnarray}
and
\begin{eqnarray}
x=x_0+x_1+...\;.
\end{eqnarray}
Let us find the coefficient at \(x_{N+\gamma-1}\) in \(bx\). It is
evident, that this coefficient is equal to
\begin{eqnarray}
\frac{2a_0c_0}{p}
\end{eqnarray}
and
\begin{eqnarray}
\{\frac{2a_0c_0}{p}\}\neq 0.
\end{eqnarray}
But the term \(ax^2\) could be represented as follows
\begin{eqnarray}
ax^2=\rm integer\; part\mit+\rm terms\;\rm which\mit\;\rm
do\;not\;depend\;of\mit\;x_{N+\gamma-1}
\end{eqnarray}
So the integral (\ref{GI}) is equal to zero.

b) So consider the case \(|a|_p>1\),
\(\left|\begin{array}{rcl}\frac{b}{a}\end{array}\right|_p\leq1\).
But each point of unit ball is its center. So, using substitution
\(x\mapsto x-\frac{b}{4a}\) we obtain:
\begin{eqnarray}
I=\chi_p \left(\begin{array}{rcl}
 -\frac{b^2}{4a} \end{array}\right) \int \limits_{B_0}\chi_p(ax^2)d\mu(x)
\end{eqnarray}
Let us represent \(a\) as follows \(a=\frac{1}{p^N}(a_0+a_1p+...)\).
At first consider the case when \(a_0\) is odd. We can represent
\(a\) as follows
\begin{eqnarray}
a=\frac{q}{p^N}d^2,
\end{eqnarray}
where \(q\) is a primer number \(p\neq q\) and \(d\) is some
\(p\)-adic number such that \(|d|_p=1\). We have
\begin{eqnarray}
I=\chi_p \left(\begin{array}{rcl}
 -\frac{b^2}{4a} \end{array}\right) \int
 \limits_{B_0}\chi_p(ax^2)d\mu(x)\nonumber\\
=\chi_p \left(\begin{array}{rcl}
 -\frac{b^2}{4a} \end{array}\right) \int
 \limits_{B_0}\chi_p(\frac{q}{p^N}x^2)d\mu(x)
\end{eqnarray}
Here we use the substitution \(x\rightarrow d^{-1}x\). We have
\begin{eqnarray}
I=\chi_p (
 -\frac{b^2}{4a}) \frac{1}{p^N}J,
\end{eqnarray}
where, by definition
\begin{eqnarray}
J=\sum \limits_{k=1}^{p^N} \rm exp \mit (\frac{2\pi i q k^2}{p^N}).
\end{eqnarray}
Let us define
\begin{eqnarray}
J_\varepsilon:=\sum \limits_{k \in \mathbb{Z}}\rm exp \mit
\{\frac{2\pi i q}{p^N}x^2-\varepsilon k^2\},
\end{eqnarray}
and
\begin{eqnarray}
f_\varepsilon:=\rm exp \mit \{\frac{2\pi i q}{p^N}x^2-\varepsilon
k^2\},\;\varepsilon>0
\end{eqnarray}
It is easy to prove
\begin{eqnarray}
J_\varepsilon=\frac{1}{p^N}J\int \limits_{-\infty}^{+\infty}\rm exp
\mit(-\varepsilon x^2)dx+O(1),\;\varepsilon\rightarrow 0.
\end{eqnarray}
Therefore
\begin{eqnarray}
J_\varepsilon=\frac{1}{p^N}
J\sqrt{{\pi}{\varepsilon}}+O(1),\;\varepsilon\rightarrow 0.
\end{eqnarray}
From other hand
\begin{eqnarray}
\hat{f}_\varepsilon\approx\sqrt{\frac{p^N}{4\pi
q}}e^{i\frac{\pi}{4}}\rm exp \mit (-\frac{ik^2 p^N}{8\pi
q}(1+\frac{p^N\varepsilon}{2\pi i q})).
\end{eqnarray}
According to Poison summarizing formula we have (if
\(\varepsilon\rightarrow 0\))
\begin{eqnarray}
J_\varepsilon=(2\pi)^{\frac{1}{2}} \sum \limits_{k
\in 2\pi\mathbb{Z}}\hat{f}_\varepsilon (k)\nonumber\\
=(2\pi)^{\frac{1}{2}}\sqrt{\frac{p^N}{4\pi q}}e^{i\frac{\pi}{4}}\sum
\limits_{k \in 2\pi \mathbb{Z}}\rm exp \mit (-\frac{ik^2 p^N}{8\pi
q}(1+\frac{p^N\varepsilon}{2\pi i q}))+O(1)\nonumber\\
=\frac{(2\pi)^{\frac{1}{2}}}{4q}\sqrt{\frac{p^N}{4\pi
q}}e^{i\frac{\pi}{4}} \sum \limits_{k=1}^{4q} \rm exp \mit
(-\frac{2\pi i k^2 p^N}{4q})\int \limits_{-\infty}^{+\infty} \rm exp
\mit
(-\frac{ k^2 p^{2N}\varepsilon}{4q^2})dk+O(1). \nonumber\\
\end{eqnarray}
We obtain from this (by calculating Gauss integral) that (if
\(\varepsilon\rightarrow 0\))
\begin{eqnarray}
J_\varepsilon=\sqrt{\frac{\pi}{\varepsilon}}\frac{1}{2\sqrt{2}}\frac{1}{\sqrt{q}}\frac{e^{i\frac{\pi}{4}}}{\sqrt{p^N}}
\sum \limits_{k=1}^{4q} \rm exp \mit (-\frac{2\pi i k^2
p^N}{4q})+O(1).
\end{eqnarray}
To calculate this sum we use the Chinese remainder theorem. Let us
define a real number \(\Sigma\) as follows
\begin{eqnarray}
\Sigma:=\sum \limits_{k=1}^{4q} \rm exp \mit (-\frac{2\pi i k^2
p^N}{4q}).
\end{eqnarray}
Since \(q\) is odd there exists such integer numbers \(r\) and \(s\)
that:
\begin{eqnarray}
4r+qs=1.
\end{eqnarray}
So by using the Chinese remainder theorem we find
\begin{eqnarray}
\Sigma=\sum \limits_{n=1}^{4} \rm exp \mit (-\frac{2\pi i n^2 s
p^N}{4}) \sum \limits_{m=1}^{q}\rm exp \mit (-\frac{2\pi i m^2 r
p^N}{q}).
\end{eqnarray}
It is easy to check that we can chose \(r\) and \(s\) such that
\(s=1\) if \(q=1 (\rm mod \mit\;4)\) and \(s=-1\) if \(q=3 (\rm mod
\mit\;4)\). In booth this cases \(r=\frac{1}{4}(\rm mod \mit\;4)\).
By using calculations performed in previous section we obtain:
\begin{eqnarray}
\Sigma=(-1)^{\frac{q-1}{2}}2\sqrt{2}\sqrt{q}e^{-i\frac{\pi}{4}}(\frac{p^N}{q})\;
\rm if \mit\; q=1(\rm mod \mit\;4)\;\rm and \mit\;\;qp^N=1(\rm mod
\mit\;4),\nonumber\\
\Sigma=(-1)^{\frac{q-1}{2}}2\sqrt{2}\sqrt{q}e^{+i\frac{\pi}{4}}(\frac{p^N}{q})\;
\rm if \mit\; q=1(\rm mod \mit\;4)\;\rm and \mit\;\;qp^N=3(\rm mod
\mit\;4),\nonumber\\
\Sigma=i(-1)^{\frac{q-1}{2}}2\sqrt{2}\sqrt{q}e^{-i\frac{\pi}{4}}(\frac{p^N}{q})\;
\rm if \mit\; q=3(\rm mod \mit\;4)\;\rm and \mit\;\;qp^N=1(\rm mod
\mit\;4),\nonumber\\
\Sigma=i(-1)^{\frac{q-1}{2}}2\sqrt{2}\sqrt{q}e^{+i\frac{\pi}{4}}(\frac{p^N}{q})\;
\rm if \mit\; q=3(\rm mod \mit\;4)\;\rm and \mit\;\;qp^N=3(\rm mod
\mit\;4).
\end{eqnarray}
If \(N\) is even we find:
\begin{eqnarray}
\Sigma=2\sqrt{2}\sqrt{q}e^{-i\frac{\pi}{4}}.
\end{eqnarray}
and
\begin{eqnarray}
I=\chi_p (
 -\frac{b^2}{4a})\frac{1}{|a|_p^{\frac{1}{2}}}
\end{eqnarray}
Now let us consider the case when \(N\) is odd. In this case
\((\frac{p^N}{q})=(\frac{p}{q})\), therefore
\begin{eqnarray}
\Sigma=(-1)^{\frac{q-1}{2}}2\sqrt{2}\sqrt{q}e^{-i\frac{\pi}{4}}(\frac{p}{q})\;
\rm if \mit\; q=1(\rm mod \mit\;4)\;\rm and \mit\;\;qp=1(\rm mod
\mit\;4),\nonumber\\
\Sigma=(-1)^{\frac{q-1}{2}}2\sqrt{2}\sqrt{q}e^{+i\frac{\pi}{4}}(\frac{p}{q})\;
\rm if \mit\; q=1(\rm mod \mit\;4)\;\rm and \mit\;\;qp=3(\rm mod
\mit\;4),\nonumber\\
\Sigma=i(-1)^{\frac{q-1}{2}}2\sqrt{2}\sqrt{q}e^{-i\frac{\pi}{4}}(\frac{p}{q})\;
\rm if \mit\; q=3(\rm mod \mit\;4)\;\rm and \mit\;\;qp=1(\rm mod
\mit\;4),\nonumber\\
\Sigma=i(-1)^{\frac{q-1}{2}}2\sqrt{2}\sqrt{q}e^{+i\frac{\pi}{4}}(\frac{p}{q})\;
\rm if \mit\; q=3(\rm mod \mit\;4)\;\rm and \mit\;\;qp=3(\rm mod
\mit\;4).
\end{eqnarray}
But, by using the Gauss reciprocity low we find:
\begin{eqnarray}
\Sigma=2\sqrt{2}\sqrt{q}e^{-i\frac{\pi}{4}}(\frac{q}{p}),\; \rm if
\mit\;p=1(\rm mod \mit\;4),\nonumber\\
\Sigma=i2\sqrt{2}\sqrt{q}e^{-i\frac{\pi}{4}}(\frac{q}{p}),\; \rm if
\mit\;p=3(\rm mod \mit\;4),
\end{eqnarray}
therefore in booth considered cases (\(N\) is even and \(N\) is odd)
we have
\begin{eqnarray}
\Sigma=2\sqrt{2}\sqrt{q}e^{-i\frac{\pi}{4}}\lambda_p(a),
\end{eqnarray}
where
\begin{eqnarray}
\lambda_p(a)=1\;\rm if\mit\;N -\rm even \mit\nonumber\\
\lambda_p(a)=(\frac{a_0}{p})\;\rm if\mit\;N -\rm odd \mit\;\rm
and\mit\;p=1(\rm mod \mit\;4),\nonumber\\
\lambda_p(a)=i(\frac{a_0}{p})\;\rm if\mit\;N -\rm odd \mit\;\rm
and\mit\;p=3(\rm mod \mit\;4).
\end{eqnarray}
So, in booth cases
\begin{eqnarray}
I=\chi_p (
 -\frac{b^2}{4a})\frac{1}{|a|_p^{\frac{1}{2}}}\lambda_p(a).
\end{eqnarray}
\begin{eqnarray}
\end{eqnarray}

Finally, let us consider the case when \(a_0\) is even. In this
case, instead of computation of
\begin{eqnarray}
I=\int \limits_{B_0}\chi_p(ax^2+bx)d\mu(x),
\end{eqnarray}
we will compute
\begin{eqnarray}
I^\star=\int \limits_{B_0}\chi_p(-ax^2-bx)d\mu(x).
\end{eqnarray}
Since \(p-a_0\) is odd we find (by using just performed
computations)
\begin{eqnarray}
I=\chi_p (
 -\frac{b^2}{4a})\frac{1}{|a|_p^{\frac{1}{2}}}\lambda'_p(a),
\end{eqnarray}
where, by definition
\begin{eqnarray}
\lambda'_p(a)=1\;\rm if\mit\;N -\rm even, \mit\nonumber\\
\lambda'_p(a)=(\frac{-a_0}{p})\;\rm if\mit\;N -\rm odd \mit\;\rm
and\mit\;p=1(\rm mod \mit\;4),\nonumber\\
\lambda'_p(a)=-i(\frac{-a_0}{p})\;\rm if\mit\;N -\rm odd \mit\;\rm
and\mit\;p=3(\rm mod \mit\;4).
\end{eqnarray}
But it is easy to prove (by using obtained in section 2 formula for
\(G.S(\frac{-1}{p})\)) that
\begin{eqnarray}
(\frac{-1}{p})=(\frac{1}{p})\;\rm
if\mit\;p=1(\rm mod \mit\;4),\nonumber\\
(\frac{-1}{p})=-(\frac{1}{p})\;\rm if\mit\;p=3(\rm mod \mit\;4).
\end{eqnarray}
So by using the (obvious) multiplicative property of Legendre
symbols we find \(\lambda'_p(a)=\lambda_p(a)\) for all integer
primers \(p\neq 2\) and for all \(a \in \mathbb{Q}_p\). In result in
the case when \(a_0\) even we find
\begin{eqnarray}
I=\chi_p (
 -\frac{b^2}{4a})\frac{1}{|a|_p^{\frac{1}{2}}}\lambda_p(a).
\end{eqnarray}
Therefore the theorem is proved.

\end{document}